\documentclass[amsmath,amssymb,aps,prb,twocolumn,superscriptaddress,10pt]{revtex4-2}
\usepackage{graphicx}
\usepackage{wasysym}
\usepackage{color}
\usepackage[normalem]{ulem}

\usepackage[colorlinks=true, linkcolor=blue, citecolor=blue, urlcolor=blue]{hyperref}
\newcommand{\glr}[1]{{\color[rgb]{0.6,0.0,0.6}  #1}}

\begin{document}

\title{Classical spin liquids in the kagome Ising model, and their link with magnetoelasticity}

\author{T. Vignau Costa}
\thanks{These authors contributed equally to this work.}
\affiliation{Instituto de Física de Líquidos y Sistemas Biológicos (IFLYSIB), UNLP-CONICET, Argentina.} 
\affiliation{Departamento de Física, Facultad de Ciencias Exactas, Universidad Nacional de La Plata, Argentina.}

\author{J. Colbois}
\thanks{These authors contributed equally to this work.}
\affiliation{Institut Néel, CNRS and Université Grenoble Alpes, 38000 Grenoble, France}

\author{G. L. Rossini}
\affiliation{Instituto de Física de Líquidos y Sistemas Biológicos (IFLYSIB), UNLP-CONICET, Argentina.} 
\affiliation{Departamento de Física, Facultad de Ciencias Exactas, Universidad Nacional de La Plata, Argentina.}

\author{F. Mila}
\affiliation{Institute of Physics, Ecole Polytechnique Fédérale de Lausanne (EPFL), CH-1015 Lausanne, Switzerland}

\author{D. C. Cabra}
\affiliation{Instituto de Física de Líquidos y Sistemas Biológicos (IFLYSIB), UNLP-CONICET, Argentina.} 
\affiliation{Departamento de Física, Facultad de Ciencias Exactas, Universidad Nacional de La Plata, Argentina.}

\author{S. A. Grigera}
\affiliation{Instituto de Física de Líquidos y Sistemas Biológicos (IFLYSIB), UNLP-CONICET, Argentina.} 
\affiliation{Departamento de Física, Facultad de Ciencias Exactas, Universidad Nacional de La Plata, Argentina.}
\affiliation{Clarendon Laboratory, Oxford University, Oxford OX1 3PU, United Kingdom}

\author{R. A. Borzi}
\affiliation{Instituto de Física de Líquidos y Sistemas Biológicos (IFLYSIB), UNLP-CONICET, Argentina.} 
\affiliation{Departamento de Física, Facultad de Ciencias Exactas, Universidad Nacional de La Plata, Argentina.}

\date{\today}

\begin{abstract}

We investigate a nearest-neighbor Ising model on the kagome lattice coupled to elastic lattice distortions. By integrating out the lattice degrees of freedom, we derive an approximate spin-only description in terms of a kagome \(J_1\)-\(J_2\)-\(J_{3}\) Ising model with interactions extending to third nearest-neighbors. Using tensor network methods complemented by classical Monte Carlo simulations, we revisit its zero-temperature phase diagram and characterize the ground states studied more than thirty years ago. We discover three 
phases with extensive zero-temperature entropy and algebraically decaying spin correlations. Importantly, these states are not confined to fine-tuned points or frontiers between phases, but occupy extended areas of parameter space. Their quasi long-range order is reflected in Bragg-like peaks in the structure factors; we relate this, and other properties, to diverse phases found in Ising and Heisenberg models. Finally, we trace the trajectories generated by the magnetoelastic coupling through the effective-model phase diagram, explaining the emergence of self-frustration for the unfrustrated magnetoelastic model with ferromagnetic interactions. 
\end{abstract}
	
\maketitle

%%%%%%%%%%%%%%%%%%%%%%%%%%%%%%%%%%
\section{Introduction}
%%%%%%%%%%%%%%%%%%%%%%%%%%%%%%%%%%

Geometrically frustrated magnetism addresses situations in which the geometry of the magnetic lattice prevents competing local interactions from being simultaneously satisfied, thus suppressing conventional long-range order~\cite{ramirez1994strongly,lacroix2011introduction,diep2013frustrated}. 
Early studies of geometrical frustration focused on two-dimensional lattices composed of triangular plaquettes of Ising spins with antiferromagnetic nearest-neighbor (NN) interactions. The triangular and kagome lattices were shown to exhibit macroscopically degenerate ground states with a finite residual entropy per site~\cite{wannier1950antiferromagnetism,syozi_1953_kagome,kano1953kagome} and decaying spin correlations~\cite{Suto_models_1981,Stephenson1970}. Subsequent studies of the kagome lattice showed that the inclusion of further-neighbor interactions gives rise to a broad variety of ground states~\cite{azaria1987coexistence,1988-Wolf,takagi1993magnetic,mizoguchi2017clustering,Colbois2021,colbois2022partial}. 

The assumption of a rigid lattice, however, is an idealization, as real magnetic materials are inevitably subjected to lattice distortions.
The coupling between lattice and magnetic degrees of freedom can qualitatively alter the magnetic behavior and stabilize phases absent in the rigid-lattice limit~\cite{Penc2004,Tchernyshyov2010}. 
Such effects are particularly relevant in frustrated systems, where structural distortions can lift magnetic degeneracies and select among competing states. 
More recently, the converse mechanism has also been demonstrated: a spontaneous lattice distortion can itself generate magnetic frustration, giving rise to an exponentially degenerate low-temperature manifold~\cite{strevcka2012spin,strevcka2019anomalous,costa2026magnetic}.  

Lattice distortions can in some cases be integrated out, yielding effective rigid-lattice models with further-neighbor magnetic interactions.
{This provides a unique route to rigid lattice models with farther couplings limited to second- or third-neighbors, as opposed to long-range dipolar couplings. Given the very different and far richer physics realized in these models with limited farther couplings, the mapping is interesting  both as effective description of magnetoelastic systems and as a means of exploring the resulting
landscape of magnetic phases.
This contrast is quite clear when applied to the kagome lattice: while dipolar couplings already lead to complex long-range orders and exotic finite-temperature phases~\cite{chern_magnetic_2012,chioar2016,canals_fragmentation_2016, hamp_supercooling_2018}, the lattice is also known for a particularly rich landscape of exotic phases arising for farther-neighbor finite-range couplings~\cite{azaria1987coexistence,1988-Wolf,takagi1993magnetic,mizoguchi2017clustering,Colbois2021,colbois2022partial,costa2026magnetic,mizoguchi_magnetic_2018-1,kiese_pinch-points_2023,lugan_schwinger_2022-3,messio2011lattice,chern_magnetic_2012}.

In this work, we investigate the magnetism derived from a simple two-dimensional magnetoelastic Ising model with nearest-neighbor (NN) interactions on the kagome lattice. Following the approach of Refs.~\cite{strevcka2012spin,strevcka2019anomalous,gen2022nematicity}, we derive an approximate effective model involving only spin degrees of freedom. The resulting magnetic system corresponds to a kagome spin model with up to third-neighbor interactions $J_1-J_2-J_{3}$, first studied by Wolf and Schotte more than three decades ago~\cite{1988-Wolf}. The kagome lattice has two geometrically distinct third-nearest-neighbor interactions. Here, the interaction at chemical distance two ($J_{3a}$) remains finite, whereas that at chemical distance three ($J_{3b}$) vanishes~\cite{mizoguchi2017clustering,Colbois2021}.

Combining classical tensor network techniques with Monte Carlo simulations, we reconstruct the zero-temperature phase diagram in the $J_3/J_1$--$J_2/J_1$ plane originally obtained by Wolf \textit{et al.}, introducing small corrections to the original paper~\cite{1988-Wolf}. Furthermore, since we have access to the ground-state degeneracy of these states, we extend the study by identifying three macroscopically degenerate phases with quasi-long-range order. We characterize and situate them in the context of other phases found in previous works on Ising~\cite{1988-Wolf,mizoguchi2017clustering,costa2026magnetic,Colbois2021,andrade_partial_2024-4} and Heisenberg~\cite{gen2022nematicity,mizoguchi_magnetic_2018-1,kiese_pinch-points_2023,lugan_schwinger_2022-3,messio2011lattice} kagome models. 
Our computations extend existing results on the thoroughly studied $J_2=J_{3}$ ($J_{3a}$) line~\cite{mizoguchi2017clustering, mizoguchi_magnetic_2018-1,lugan_schwinger_2022-3,kiese_pinch-points_2023}.

We then return to the original magnetoelastic model and trace how variations in its parameters define trajectories through the phase diagram of the effective spin model. This provides further insight into its phase behavior and clarifies why self-frustration emerges in the ferromagnetic kagome lattice but is absent in the antiferromagnetic case~\cite{costa2026magnetic}.

The paper is organized as follows. In Sect.~\ref{sec:Model&Methods}, we introduce the Einstein site-phonon model and show that integrating out the elastic degrees of freedom yields effective spin Hamiltonians with further-neighbor interactions.
In Sect.~\ref{sec:Numerical_Methods} we describe the tensor-network formalism employed to determine the residual entropy and ground state spin-spin correlations of the effective models. 
In Sect.~\ref{sec:Numerical_Results} we reconstruct and partially revise the magnetic phase diagram of the kagome lattice originally proposed by Wolf \textit{et al.}~\cite{1988-Wolf}. We extend the scope of that work by identifying  different disordered phases and determining their residual entropies and spin structure factor. A preliminary version of the residual entropy results presented here using tensor networks was previously discussed in the PhD thesis of one of the authors \cite{2022-Colbois}; the present work extends these results, includes the spin correlations and structure factors, and places them in the context of magnetoelastic effects.
In Sect.~\ref{sec:back_to_magnetoelatics} we turn back to the connection between the magnetic effective and magnetoelastic models, focusing on the origin of the phenomenon of self-frustration introduced in Ref.~\onlinecite{costa2026magnetic}. 
The implications of these results are discussed in Sect.~\ref{sec:dis}. Finally, our conclusions are presented in Sec.~\ref{sec:sum_conc}.

%%%%%%%%%%%%%%%%%%%%%%%%%%%%%%%%%%
\section{\glr{Effective magnetic spin models from Magnetoelastic coupling}} \label{sec:Model&Methods}
%%%%%%%%%%%%%%%%%%%%%%%%%%%%%%%%%%

Families of purely magnetic spin models with further-neighbor interactions can arise from simple nearest-neighbor models with magnetoelastic coupling. In order to show this, we consider a lattice of magnetic ions carrying Ising spins and allow the ions to undergo elastic displacements from their equilibrium positions. The magnetoelastic coupling is described within the framework of the \emph{Einstein site-phonon spin model} \cite{wang2008app}, in which the lattice distortions are modeled as  independent harmonic oscillators. Despite its simplicity, this model captures the essential contribution of optical phonons to lattice distortions when the magnetic lattice is embedded in a three-dimensional crystal. The resulting system is described by the following dimensionless Hamiltonian:
\begin{equation}
\frac{\mathcal{H}(\{\sigma\},\{\delta\mathbf{r}\})}{|J_0|}=\sum_{{\langle i,j \rangle}_1} j_{ij}(\textbf{r}_{ij}) \sigma_i \sigma_j  +\frac{K_e}{2} \sum_i (\delta \mathbf{r}_{i})^2 
\label{eq:H}
\end{equation}
where $\sigma_i=\pm1$ denotes an Ising spin at site $i$, and $\langle i,j\rangle_1$ indicates nearest-neighbor pairs. The position of spin $i$ is given by $\mathbf{r}_i$, expressed in units of the lattice constant.
The site displacement is defined as $\delta\mathbf{r}_i\equiv\mathbf{r}_i-\mathbf{r}_i^0$, where $\mathbf{r}_i^0$ is the equilibrium position of site $i$.
The dimensionless nearest-neighbor exchange couplings $j_{ij}$ depend on the relative
position vectors between neighboring spins denoted as $\mathbf{r}_{ij}=\mathbf{r}_i-\mathbf{r}_j$.  Here, $J_0$ is the nearest-neighbor exchange energy of the undistorted lattice, and $K_e$ is a dimensionless elastic constant chosen to be sufficiently large that the crystal melting temperature, estimated according to the Lindemann criterion, is separated from the Curie temperature (set by $J_0$) by several orders of magnitude~\cite{pili2019two}. 

We assume that the position dependence of exchange couplings $j_{ij}$ can be expanded around their undistorted values
as a linear function of the distance between the spins $r_{ij}$, 
\begin{equation}
j_{ij}(\textbf{r}_{ij})=\mathrm{sgn}(J_0)\,[1-\alpha ( r_{ij}-1 )]
    \label{eq:distance_linear}
\end{equation}
where $r_{ij}=|\textbf{r}_{ij}|$ and $\alpha>0$ is the magnetoelastic coupling constant. 

The derivation of the effective spin Hamiltonian begins by further expanding $r_{ij}$ in terms of the ion displacements $\delta\textbf{r}_{i}$ as 
$r_{ij} = 
|\breve{\boldsymbol{e}}_{ij}+\delta\textbf{r}_{j}-\delta\textbf{r}_{ji}|
\approx 
1 + \breve{\boldsymbol{e}}_{ij}\cdot(\delta\textbf{r}_{j}-\delta\textbf{r}_{i})$ where $\breve{\mathbf{e}}_{j}$ is the unit vector pointing from site $i$ to site $j$. The exchange couplings to linear order in the limit of small ionic displacements are then
\begin{equation}
    j_{ij}\approx\mathrm{sgn}(J_0)\,[1-\alpha\ \breve{\boldsymbol{e}}_{ij}\cdot(\delta\textbf{r}_{j}-\delta\textbf{r}_{i})]
    \label{eq:linearization}\ .
\end{equation}

Substituting this expansion into Eq.~\eqref{eq:H}, the Hamiltonian can be rewritten as
\begin{equation}
\begin{aligned}
    \frac{\mathcal{H}'(\{\sigma\},\{\delta\mathbf{r}\})}{|J_0|}& = \sum_{\langle i,j \rangle_1}\mathrm{sgn}(J_0)\sigma_i\sigma_j + \\
    & \sum_i\left[\frac{K_e}{2}(\delta \mathbf{r}_{i})^2-\bigg[\sum_{j\in n(i)}\mathbf{F}_{ij}\bigg]\cdot\delta \mathbf{r}_{i}\right]
\end{aligned}
\end{equation}
where $n(i)$ is the set of nearest-neighbors of site $i$.  We have introduced the dimensionless magnetic force
\begin{equation}
    \mathbf{F}_{ij} = -\mathrm{sgn}(J_0)\ \alpha\ \sigma_i\sigma_j\ \breve{\boldsymbol{e}}_{ij},
\end{equation}
which acts between nearest-neighbor sites and depends on the magnetic configuration and the bond orientation.

Completing the square, the approximate Hamiltonian becomes
\begin{equation}
\begin{aligned}
    \frac{\mathcal{H}'}{|J_0|}& = \sum_{\langle i,j \rangle}\mathrm{sgn}(J_0)\sigma_i\sigma_j - \frac{1}{2K_e}\sum_i\left[\sum_{j,l\in n(i)}\mathbf{F}_{ij}\cdot\mathbf{F}_{il}\right] \\
    & +\frac{K_e}{2}\sum_i\left[\delta \mathbf{r}_{i}-\frac{1}{K_e}\sum_{j\in n(i)}\mathbf{F}_{ij}\right]^2.
    \label{eq:H_completedsquares}
\end{aligned}
\end{equation}

A Gaussian integration of the last term over elastic thermal
fluctuations gives a trivial contribution to the partition function (an energy of $1k_B T$ per spin, in accordance with the equipartition of energy), leaving an effective spin-only Hamiltonian.

In the resulting effective spin model 
\begin{equation}
    \mathbf{F}_{ij}\cdot\mathbf{F}_{il} = \alpha^2\ \breve{\boldsymbol{e}}_{ij}\cdot\breve{\boldsymbol{e}}_{il}\ \sigma_j\sigma_l
\end{equation}
can be interpreted as an effective exchange interaction between spins $\sigma_j$ and $\sigma_l$, mediated by their common nearest neighbor $\sigma_i$. Such a contribution arises only when both sites $j$ and $l$ are nearest neighbors of the same site $i$. Its magnitude and sign are determined by the bond geometry through the scalar product $\breve{\mathbf{e}}_{ij}\cdot\breve{\mathbf{e}}_{il}$. 

Subtracting the trivial contribution from equipartition, the spin-only effective Hamiltonian therefore takes the form
\begin{equation}
    \frac{\mathcal{H}_{\rm eff}(\{\sigma\})}{|J_0|} = \sum_{\langle i,j \rangle_1}\mathrm{sgn}(J_0)~\!\sigma_i\sigma_j-\sum_{\langle j\neq l\rangle_i} \frac{\alpha^2}{2K_e}(\breve{\boldsymbol{e}}_{ij}\cdot\breve{\boldsymbol{e}}_{il})~\sigma_j\sigma_l
\end{equation}
where the second sum runs over pairs of sites that share a common neighbor $i$. We have also omitted a constant term that arises from the contributions with $j=l$.

%%%%%%%%%%%%%%%%%%%%%%%%%%%%%%%%%%

%%%%%%%%%%%%%
\begin{figure}
\begin{centering}
\includegraphics[scale=0.31,angle=0]{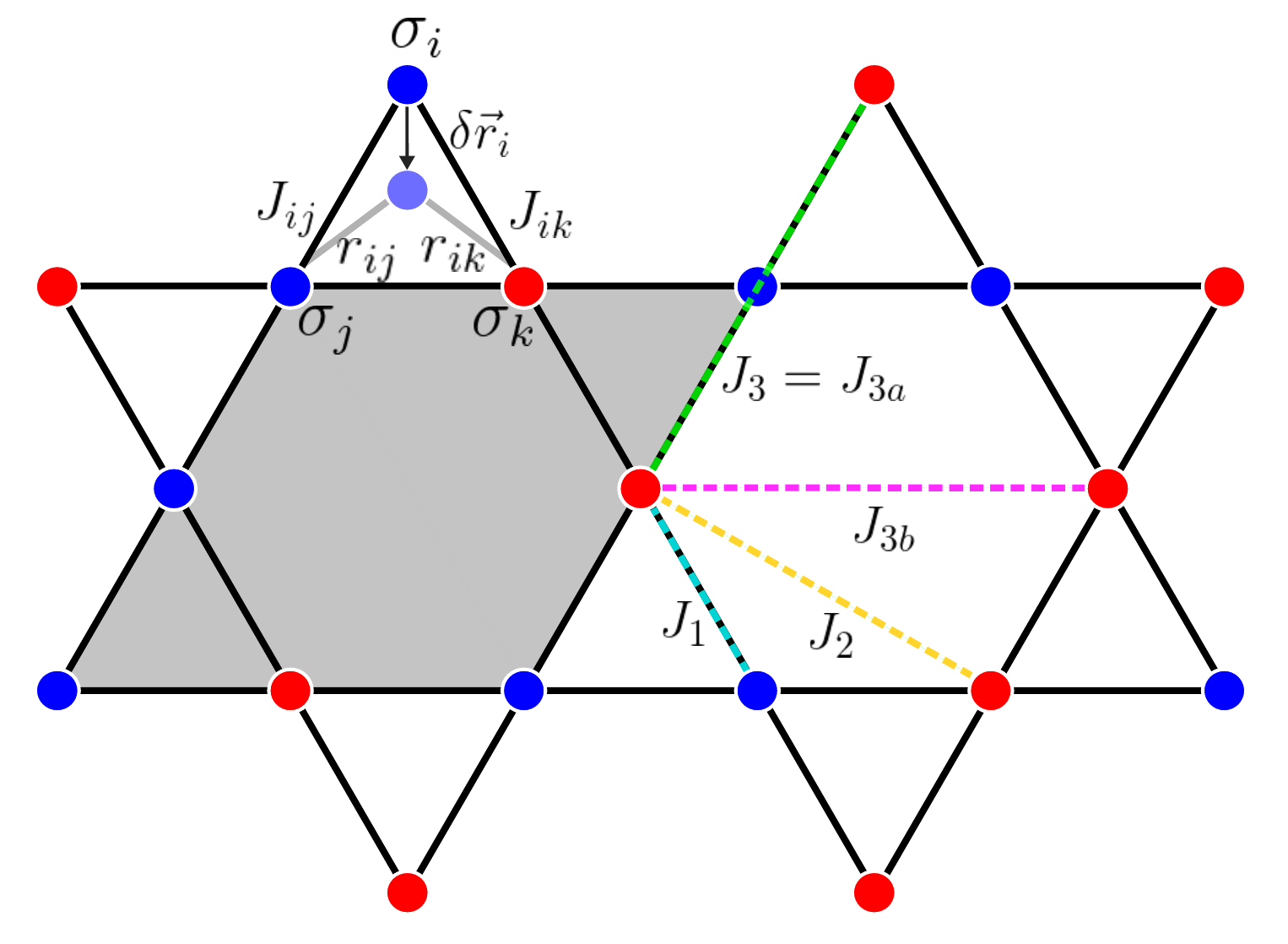}
\par\end{centering}
\caption{Kagome lattice geometry studied on this work. A distortion at site $i$ is indicated together with its effect on the neighboring
exchange couplings and nearest-neighbors distances. Circles represent Ising spins with positive (blue) and negative (red) orientation along the magnetic axis. The arrow indicates an ion displacement from the pristine lattice (bold dots) to its new position on the lattice (translucent dot). We represent with a dotted line the distances of first (light blue), second (yellow), and third (green, magenta) neighbors. We distinguish the two different third neighbors as being bridged by an intermediate site (green) or not (magenta). The unit cell is painted in gray.
}
\label{fig:lattices}
\end{figure}
%%%%%%%%%%%%%

We now turn to the particular case under study: the kagome lattice. Every nearest-neighbor pair shares a common neighboring site (see Fig.~\ref{fig:lattices}), with $\breve{\mathbf{e}}_{ij}\cdot\breve{\mathbf{e}}_{il}>0$, while every second-nearest-neighbor pair is connected through an intermediate site for which $\breve{\mathbf{e}}_{ij}\cdot\breve{\mathbf{e}}_{il}<0$. The situation is slightly different for third neighbors: of the six third-neighbor pairs associated with a given site, four are connected through a common intermediate site, whereas the remaining two---located at opposite vertices of adjacent hexagons---are not. This gives rise to two inequivalent third-neighbor couplings, denoted by $J_{3a}$ and $J_{3b}$ in Fig.~\ref{fig:lattices}. Collecting these contributions yields
\begin{align}
J_1 &= J_0-\frac{\alpha^2|J_0|}{2K_e}, &
J_2 &= \frac{\alpha^2|J_0|}{2K_e}, \nonumber\\
J_{3a} &= \frac{\alpha^2|J_0|}{K_e}\equiv J_3, &
J_{3b} &= 0 .
\label{eq:kagome}
\end{align}
Thus, the magnetoelastic correction to $J_1$ is always ferromagnetic, whereas the induced $J_2$ and $J_3$ interactions are always antiferromagnetic, independently of the sign of the bare exchange $J_0$. In the ferromagnetic case the correction to $J_1$ reinforces the bare interaction, while in the antiferromagnetic case it competes with it and, for sufficiently large $\alpha$, eventually changes its sign. 

The resulting effective Hamiltonian is
\begin{equation}
    \mathcal{H}_{\rm eff} = \sum_{\langle i,j \rangle_1} J_1~\sigma_i\sigma_j+\sum_{{\langle  i,j \rangle}_2} J_2~\sigma_i\sigma_j+\sum_{{\langle  i,j  \rangle}_3} J_3~\sigma_i\sigma_j 
    \label{eq:Heff}
\end{equation}
where $\langle i,j\rangle_n$ denotes pairs of $n$th-neighbor sites, with the third-neighbor sum restricted to pairs at chemical distance two.

\section{Numerical methods \label{sec:Numerical_Methods}}
%%%%%%%%%%%%%%%%%%%%%%%%%%%%%%%%%%

To determine the complete ground-state phase diagram of the kagome lattice model and evaluate the residual entropy of its macroscopically degenerate phases, we employ classical Monte Carlo (MC) simulations and tensor-network (TN) methods as complementary approaches.

%%%%%%%%%%%%%%%%%%%%%%%%%%%%%%%%%%
% subsection{Tensor Networks}  % there is no other subsection

While tensor networks are most commonly associated with quantum many-body systems, their historical development is closely connected to transfer-matrix techniques~\cite{KramersWannier1941,Baxter1968,Baxter1978,NishinoDMRGList,Okunishi2022}, and they provide an equally natural framework for studying Ising models on lattices~\cite{Nishino1996,Levin2007,Orus2008}, albeit with some important subtleties in the case of frustrated models~\cite{Vanderstraeten2018, Vanhecke2021, colbois2022partial,Song2023}.

The central idea of using TNs for two-dimensional translation-invariant statistical mechanics models on a lattice is to represent the partition function as the evaluation of a two-dimensional tensor network. This is done in two steps. First, we take advantage of translation invariance by encoding the partition function on the infinite 2D lattice using a repeated local tensor. For Ising models, this encoding is exact. Second, the contraction (i.e. the evaluation) of the resulting infinite 2D tensor network is performed \emph{approximately} using standard methods~\cite{Levin2007,Nishino1996,Orus2009,ZaunerStauber2018,Vanderstraeten2019_Tangent, vanthilt_practical_2026}.

\subsection{Construction of the local tensor}

A given partition function admits several different encodings as a 2D TN. However, in the case of frustrated models, the convergence of the second step (the approximate contraction) depends on the choice of encoding~\cite{Vanhecke2021}. For constructing the tensor network, we follow the method reported in Refs.~\onlinecite{Vanhecke2021,colbois2022partial, 2022-Colbois}. We provide a brief reminder and justification for the general strategy in Appendix~\ref{app:TN}. The code for the tensor network construction and contraction will be made available upon publication. 
%%%%%%%%%%%%%
\begin{figure}
\begin{centering}
\includegraphics[scale=0.7,angle=0]{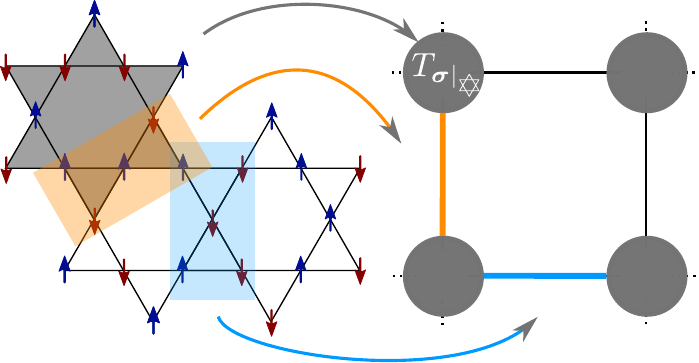}
\par\end{centering}
\caption{Spirit of the tensor network construction. Left: a spin configuration on a patch of the kagome lattice, seen as a set of overlapping star tiles. Right: a patch of the 2D infinite tensor network representing the partition function, which is a weighted sum over all possible configurations. Each kagome star maps to a tensor in the tensor network construction shown on the right, see Eq.~\eqref{eq:tensor}. The contraction of neighboring tensors automatically enforces the matching conditions for overlapping stars, as highlighted by the blue and orange rectangles. 
}
\label{fig:tensor}
\end{figure}
%%%%%%%%%%%%%

The general idea is to factorize the Boltzmann weight of any given configuration over a set of overlapping clusters, with each tensor encoding the local contribution associated with a cluster. Contracting the tensor indices then sums over all configurations while enforcing the appropriate matching conditions. This construction is illustrated schematically in Fig.~\ref{fig:tensor}

For the effective Hamiltonian in Eq. ~\eqref{eq:Heff}  it is sufficient to take a single kagome star as the elementary cluster. The kagome lattice can then be viewed as a two-dimensional triangular lattice of overlapping kagome stars, with each pair of corner-sharing triangles shared by two neighboring stars. To each kagome star, we associate a local tensor
\begin{equation}
\label{eq:tensor}
T_{\boldsymbol{\sigma}|_{\davidsstar}} = \exp\left(-\beta (h(\boldsymbol{\sigma}|_{\davidsstar})-h_0\right).
\end{equation}
Here, $\boldsymbol{\sigma}|_{\davidsstar}$ is a spin configuration on the cluster,  $h_0$ denotes the ground-state energy per cluster, and the local Hamiltonian $h$ is defined such that for any global spin configuration $\boldsymbol{\sigma}$, $H(\boldsymbol{\sigma}) = \sum_{\davidsstar} h(\boldsymbol{\sigma}|_{\davidsstar})$. The specific construction of the local Hamiltonian $h$ to ensure convergence  for a given set of couplings is discussed in Appendix~\ref{app:TN}, but the essential aspect to retain is that couplings are split between neighboring clusters in such a way that all $h$ can be simultaneously minimized~\cite{Huang2016,Vanhecke2021}, i.e., the frustration is relaxed~\footnote{Note that this is the generalization to the $J_1-J_2-J_{3}$ model of the proof of the triangular lattice Ising model ground state, whereby the Hamiltonian is rewritten as a sum over triangles of 3-spin hamiltonians which are minimized by the 2-up 1-down, 2-down 1-up configurations~\cite{Vanhecke2021,Huang2016}}. 

Two local configurations $\boldsymbol{\sigma}|_{\davidsstar}$ on two neighboring kagome stars match if they agree on the values of the spins in the overlapping region. In the tensor network, this matching condition is automatically enforced by the contraction of the legs corresponding to these sites. 
Thus, the contraction of the local tensor in Eq.~\eqref{eq:tensor} corresponds to the partition function
\begin{equation}~
    \mathcal{Z} = \sum_{\boldsymbol{\sigma}} \exp\left(-\beta(H(\boldsymbol{\sigma})-E_0)\right)
\end{equation}
where the contribution of the ground state energy has been factorized out. This construction ensures that in the low temperature limit, $\mathcal{Z}$ tends to the number of ground states and that
\begin{equation}
\label{eq:tensor2}
T_{\boldsymbol{\sigma}|_{\davidsstar}} \rightarrow  \begin{cases}
        1 \quad \text{if } \boldsymbol{\sigma}|_{\davidsstar} \text{ belongs to the ground state}\\
        0 \quad \text{otherwise}
    \end{cases}.
\end{equation}

\subsection{Evaluation}

In general, the contraction of the tensor network required to evaluate the partition function cannot be performed exactly. We therefore approximate it using the corner-transfer-matrix renormalization group (CTMRG) method~\cite{Nishino1996,Orus2008,brehmer_pepskit_2026,PEPSKitZenodo}. 

To this end, we first take advantage of the redundancy in the matching conditions to map the triangular lattice tensor network of edge-sharing stars onto an equivalent tensor network defined on the square lattice, and directly employ standard CTMRG contraction on the square lattice.

CTMRG consists in approximating the infinite environment of a local tensor using a finite set of corner and edge tensors with environment bond dimension $\chi$. 
At each step, the local tensor is absorbed in the environment, updating the corner and edge tensors; then, the enlarged environment is truncated back to the finite bond dimension $\chi$. Since the local tensor breaks the square lattice symmetries, we use the general CTMRG environment without lattice symmetries~\footnote{Importantly, in our simulations the initial environment is constructed from the local tensor by summing over one and two open legs for the edge and corner tensors, respectively. We use PEPSKit.jl v.0.8.1~\cite{PEPSKitZenodo} , growing the environment and using the 'full infinite' projectors and the 'simultaneous CTMRG' absorption step.}.

%%%%%%%%%%%%%%%
\begin{figure*}
\begin{centering}
\includegraphics[width= \textwidth]{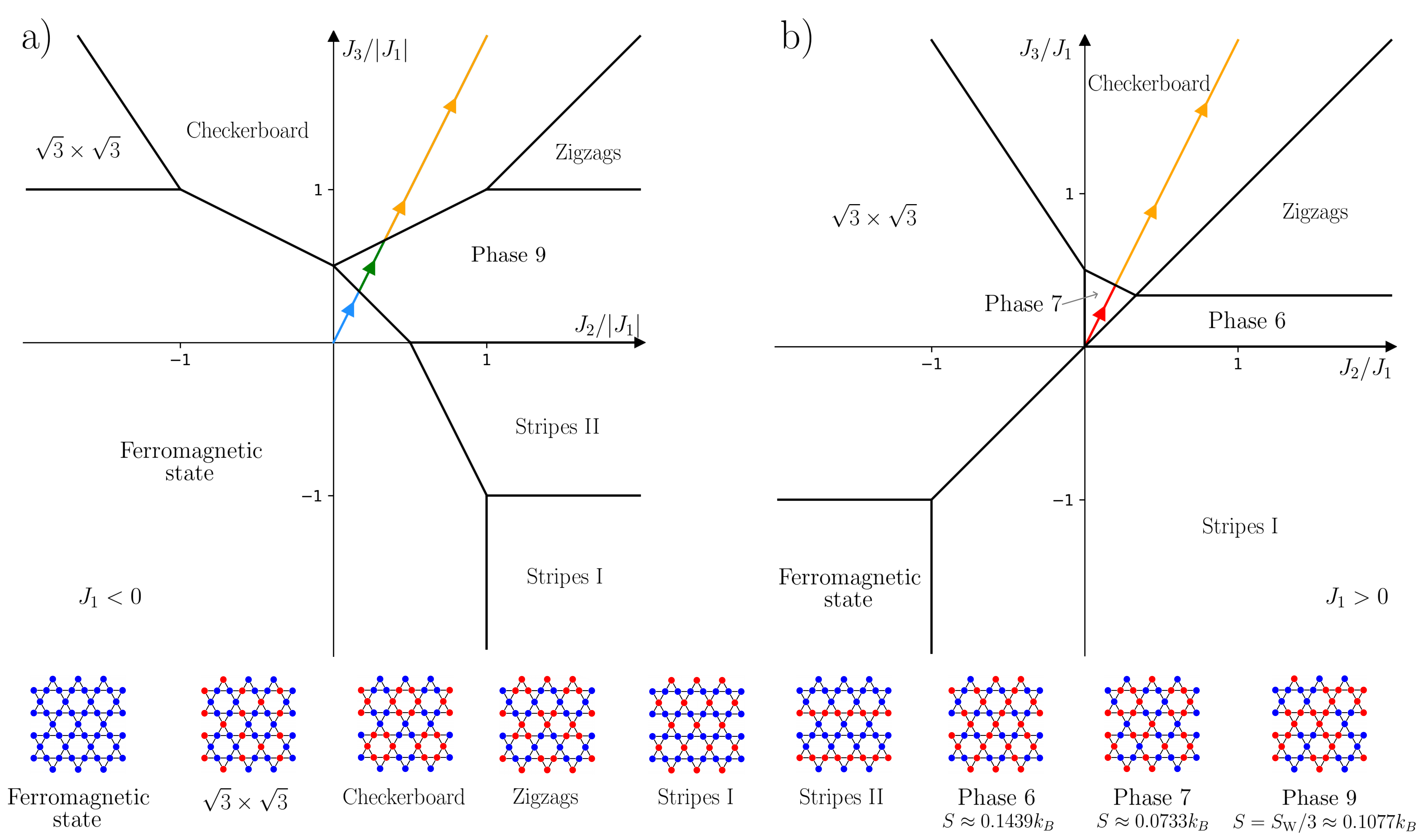}
\par\end{centering}
\caption{{\em Ground-state phase diagrams for the kagome lattice} with $J_1<0$ (left) and $J_1>0$ (right). The overall topology of the phase diagrams agrees with that reported by Wolf and Schotte~\cite{1988-Wolf}. Labels are assigned in reference to ordered magnetic patterns (noticing that $\sqrt{3}\times \sqrt{3}$ and Stripes I were mislabeled in the referred work).
Representative spin configurations are shown at the bottom of the figure, where blue and red circles denote up and down spins.
Disordered phases are labeled by their original numbers, noticing  their zero temperature residual entropies. 
The colored curves indicate the trajectories traced by the effective couplings generated from the magnetoelastic Hamiltonian. Phase 9 and Phase 7 correspond to the ``self-frustrated'' and ``paramagnetic phases'' identified in simulations of magnetoelastic models (\cite{pili2019two,costa2026magnetic}).}
\label{fig:PDkagome}
\end{figure*}
%%%%%%%%%%%%%%%

The environment bond dimension is the main parameter controlling the validity of the approximation. For partition functions with exponentially decaying correlations or long range order, the convergence is exponential in the bond dimension; whereas for power-law decay of correlations,  as we have here for certain phases, the finite bond dimension is a relevant perturbation which introduces a finite correlation length. 

At zero temperature the residual entropy per site is obtained from the tensor network according to $ S = \ln(\lambda) / 3 $, where the factor $3$ comes from the number of sites per unit cell of the local tensor, and where $\lambda$ is the leading eigenvalue per tensor of the transfer matrix for a given environment.  The convergence of the residual entropy with increasing bond dimensions for the three macroscopically degenerate phases found for the kagome lattice is discussed in the following section. The residual entropy results have been verified with an alternative tensor network contraction (variational uniform matrix product states~\cite{ZaunerStauber2018,Vanderstraeten2019_Tangent}).

\subsection{Constructing the ground state phase diagram}
The phase diagram in Fig.~\ref{fig:PDkagome} is verified using the method of inequalities for the correlation functions employed by Wolf and Schotte ~\cite{kanamori1966magnetization,brandt1986ground,1988-Wolf}, which provides a rigorous proof of the ground state energy.  It is additionally confirmed by the construction of the local ground states $\boldsymbol{\sigma}|_{\davidsstar}^{(0)}$ needed to build the tensor network~\cite{Vanhecke2021,Huang2016}. Importantly, within one phase, the local ground state configurations $\boldsymbol{\sigma}|_{\davidsstar}^{(0)}$ are independent of the value of the couplings~\cite{Vanhecke2021}. As a result, at zero temperature, the residual entropies and correlations are also independent of the couplings within each phase. Thus, for each phase, only one set of couplings needs to be studied. For phases where the number of local ground state configurations is too large to confirm by hand whether there is magnetic long-range order, a tensor network contraction is used to check whether they retain a finite residual entropy per site.

%%%%%%%%%%%%%%%%%%%%%%%%%%%%%%%%%%
\section{Results} \label{sec:Numerical_Results}
%%%%%%%%%%%%%%%%%%%%%%%%%%%%%%%%%%

%%%%%%%%%%%%%%%%%%%%%%%%%%%%%%%%%%
\subsection{Magnetic phases}\label{subsec:Res_Kag}

Figure~\ref{fig:PDkagome} shows the zero-temperature phase diagram of the effective $J_1-J_2-J_3$ Ising model. The left and right panels correspond to ferromagnetic ($J_1<0$) and antiferromagnetic ($J_1>0$) nearest-neighbor interactions, respectively. The overall topology agrees with the phase diagrams reported by Wolf and Schotte~\cite{1988-Wolf}. We note, however, that the labels of phases 2 and 3 in Ref.~\onlinecite{1988-Wolf} appear to have been inadvertently interchanged relative to their descriptions in the text, in both the ferromagnetic and antiferromagnetic cases. Representative spin configurations for all ground states are shown below the corresponding phase diagrams, with blue and red circles denoting up and down spins, respectively.

When both $J_2$ and $J_3$ are large and dominate over $J_1$, the ground states are ordered and, as expected, independent of the sign of $J_1$. We label these phases according to their characteristic spin patterns. The ferromagnetic phase is stabilized for large negative $J_2$ and $J_3$, while when both couplings are positive, the system forms either the Zigzags or Checkerboard (named after Refs.\cite{pili2019two,costa2026magnetic}) phase, depending on whether $J_2$ or $J_3$ dominates.  Large positive $J_3$ and negative $J_2$ result in the $\sqrt{3}\times\sqrt{3}$, while when signs are reversed, the system stabilizes the Stripes I phase. 

As $J_2$ and $J_3$ are reduced, the phase diagram becomes richer, with four additional phases, two for each sign of $J_1$. One of these, Stripes II, is ordered and corresponds to a variation of the Stripes I pattern that gains a small energetic advantage from a ferromagnetic $J_1$. Our calculations show that the remaining three phases, all located in the first quadrant, keep a residual entropy at zero $T$: Phase 9 for ferromagnetic $J_1$ and phases 6 and 7 (retaining Wolf and Schotte's nomenclature in these three cases~\cite{1988-Wolf}) for $J_1>0$. Their zero-temperature entropy per site, discussed below, are indicated in Fig.~\ref{fig:PDkagome}.

\subsubsection{Ferromagnetic NN interactions ($J_1<0$)}

In the first quadrant, increasing $J_2$ and $J_3$ drives a transition from the ferromagnetic state into Phase 9. Low-temperature Monte Carlo simulations reveal an extensive manifold of configurations in which the symmetry between the ``up'' and ``down'' triangles of the kagome lattice is spontaneously broken. In each configuration, one set of triangles hosts three ferromagnetically aligned spins, while the complementary set hosts configurations with two spins pointing in one direction and the third in the opposite direction, as illustrated schematically in Fig.~\ref{fig:PDkagome}. Because neighboring triangles share vertices, these local constraints do not uniquely determine the spin configuration but instead give rise to an extensively degenerate manifold of low-temperature states.

We determined the residual entropy of Phase 9 using two independent approaches: tensor networks and Monte Carlo simulations.  Figure~\ref{fig:EntropySF}a) shows the specific heat for this phase (green curve) obtained from Monte Carlo simulations on the magnetic model; the sharp peak in $C_V$ marks the spontaneous symmetry breaking transition into the phase that distinguishes between both sets of triangles. In panel b) we can see the entropy of this phase (green curve), calculated through the corresponding thermodynamic integration. Assuming zero residual entropy leads to a high-temperature entropy that falls short of the expected value of $\ln 2$. The addition of a zero-point entropy per spin of $0.1076897$, found using tensor networks (see Fig.~\ref{fig:convS}) rectifies this situation.  This value for $s_{\mathrm{res}}$ coincides within errors with $s_{\rm Wannier}/3$, with $s_{\rm Wannier}$ the Wannier entropy of the antiferromagnetic Ising model on the triangular lattice~\cite{wannier1950antiferromagnetism}. This can be readily understood: the elementary building blocks of Phase 9 are triangles of uniformly aligned spins, whose centers form a triangular lattice (see Fig.~\ref{fig:PDkagome}). Consequently, there is one effective degree of freedom for every three spins. In fact, there is an exact mapping on the triangular lattice Ising antiferromagnet's ground state, and the residual entropies must be the same to a factor three. As will be discussed below, within the framework of the magnetoelastic model, Phase 9 corresponds to the Self Frustrated phase found in Ref.~\onlinecite{costa2026magnetic}.

%%%%%%%%%%%%%%%%
\begin{figure}
\begin{centering}
\includegraphics[scale=0.14]{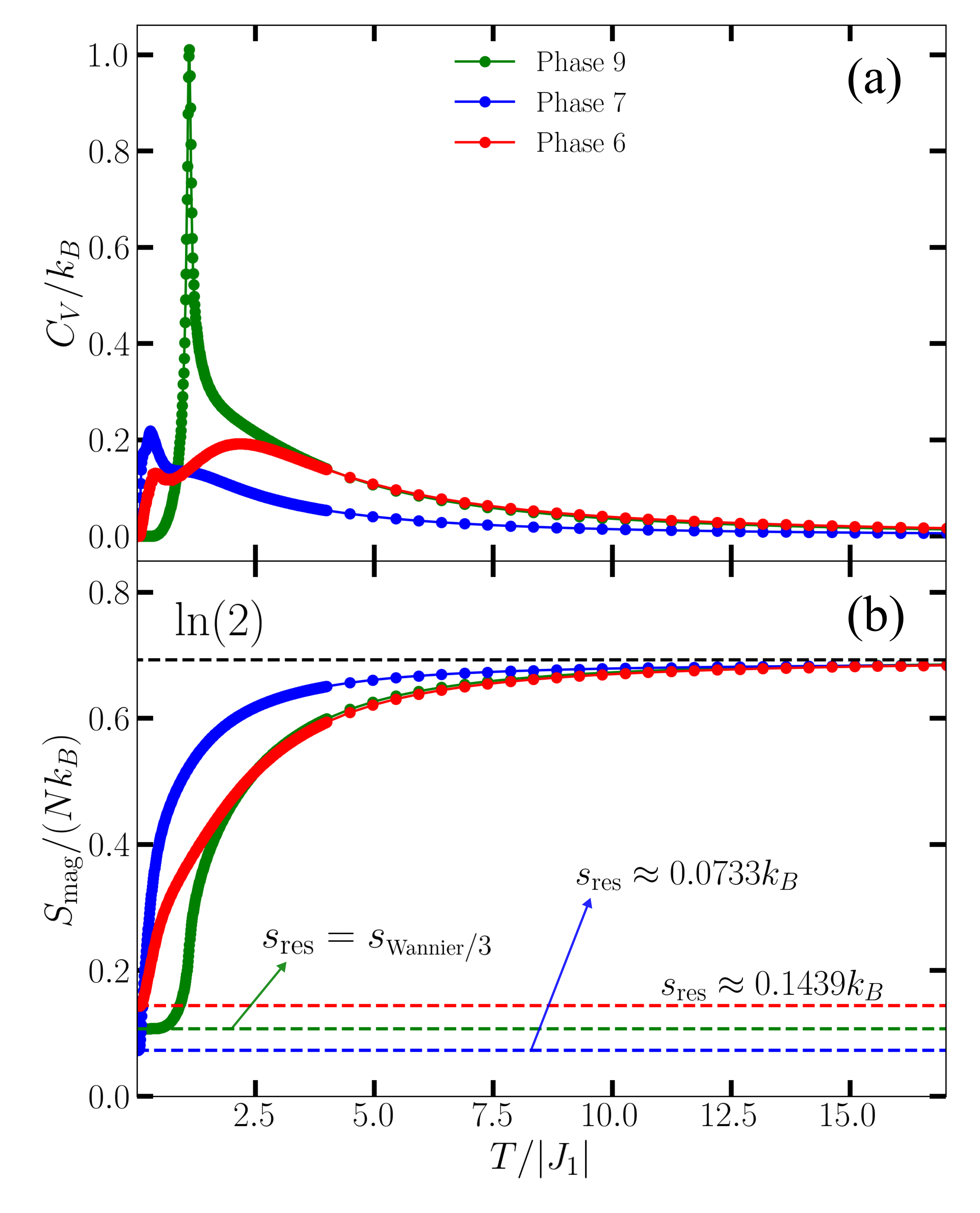}
\par\end{centering}
\caption{Specific heat (panel a)) and entropy (b)) per spin as functions of temperature for the macroscopically degenerate phases (Phase 6, 7, and 9). In each case the entropy approaches a non zero residual value indicated on the plot. The narrow peak in $C_V$ for Phase 9 shows strong finite size effects; they indicate the symmetry breaking transition between up and down triangles. Much smoother features are observed for phases $6$ and $7$, with no indications of a second order phase transition (see Appendix). The colored dashed lines in panel b) indicate the zero-point entropy added to the integrated specific heat, so that the entropies reach the value of $\log(2)$ at high $T$. The data were obtained from Monte Carlo simulations on a lattice of linear size $L=24$. The values of $J_i$ used in each case are: $J_1=1$, $J_2=1.5$, $J_3=0.15$ (Phase 6); $J_1=1$, $J_2=0.1$, $J_3=0.25$ (Phase 7); $J_1=-1$, $J_2=1$, $J_3=0.5$ (Phase 9).}
\label{fig:EntropySF}
\end{figure}
%%%%%%%%%%%%%%%%

%%%%%%%%%%%%%%%%
\begin{figure*}
\begin{centering}
\includegraphics[width = \textwidth]{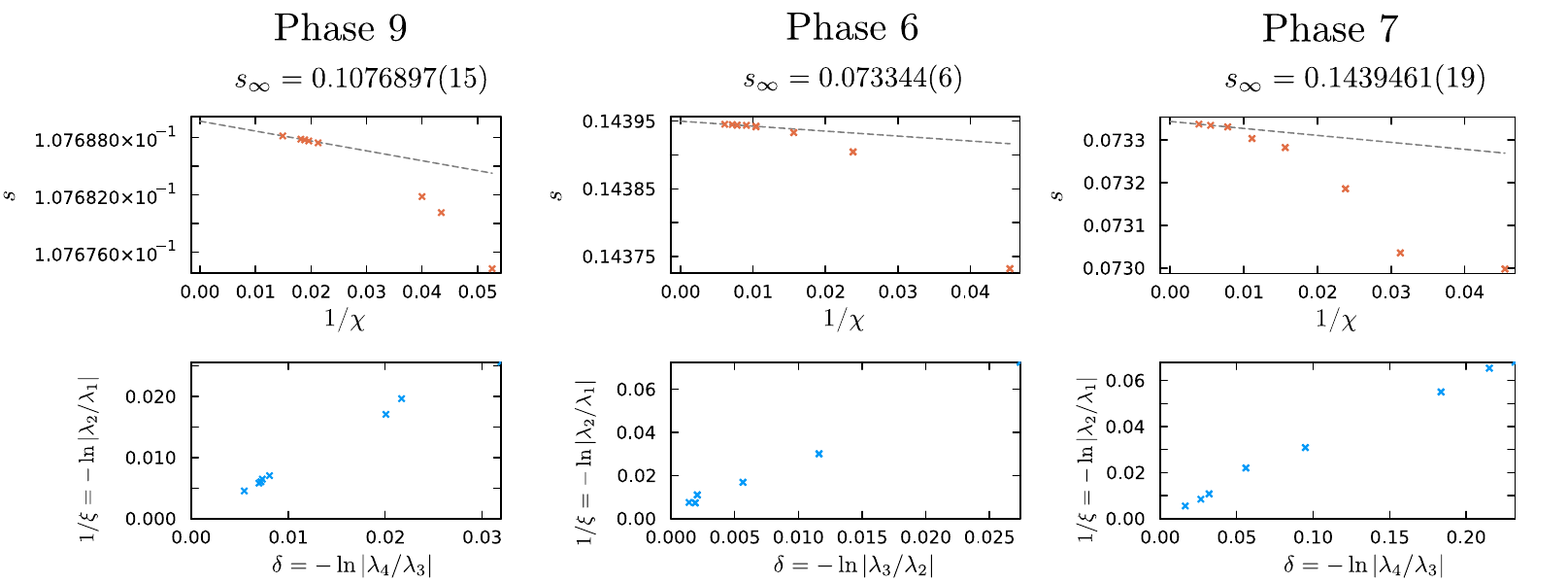}
\par\end{centering}
\caption{Residual entropy (top row) and convergence of the largest inverse correlation length (bottom row) for the three macroscopically degenerate phases (see also ~\cite{2022-Colbois, Colbois2021} for the residual entropies). On both rows, the increasing bond dimensions are from right to left. The top row shows the residual entropy as a function of the inverse bond dimension. The bottom row shows the inverse correlation length as a function of a characteristic length corresponding to the next non-degenerate gap in the transfer matrix spectrum. The values of the couplings used in each case are  $J_1=-1$, $J_2=0.5$, $J_3=0.5$ (Phase 9); $J_1=1$, $J_2=0.5$, $J_3=0.1$ (Phase 6); $J_1=1$, $J_2=0.125$, $J_3=0.25$ (Phase 7).}
\label{fig:convS}
\end{figure*}
%%%%%%%%%%%%%%%%

\subsubsection{Antiferromagnetic NN interactions ($J_1 >0$)}

In the first quadrant of the antiferromagnetic phase diagram, phases 6 and 7 appear at relatively small $J_3$. Both tensor-network calculations and Monte Carlo simulations indicate that their ground states are macroscopically degenerate. Figure~\ref{fig:EntropySF}(a) shows the specific heat from Monte Carlo as a function of temperature for these phases (blue and red curves). Although both exhibit peculiar features at low temperature, these do not show the significant finite-size dependence characteristic of a thermodynamic phase transition (see Appendix \ref{app:finite_size_effects}). This behavior contrasts with the sharp peak observed for Phase 9 (green curve).

The corresponding entropy curves, obtained by integrating $C_V/T$, are shown in Fig.~\ref{fig:EntropySF}(b). Adding the zero-temperature entropies per spin, $s_0/k_B=0.1439$ for Phase 6 and $s_0/k_B=0.0733$ for Phase 7, yields the expected high-temperature limit $s/k_B=\ln 2$. These zero-temperature entropies coincide with what is obtained from tensor networks: a linear extrapolation with increasing bond dimension, shown at the top row of Fig.~\ref{fig:convS}, provides an upper estimate, while the entropy obtained at the largest accessible bond dimension provides a lower one. The values quoted above and shown in the figure account for the uncertainty associated with these two bounds. In particular, the residual entropy of phase 9 matches $s_{\mathrm{Wannier}/3}$ within errors.

Preliminary further insight into the nature of these phases can be obtained from the transfer-matrix spectrum. The bottom row of Fig.~\ref{fig:convS} shows the inverse of the longest correlation length associated with connected correlations,
\begin{equation}
\xi^{-1}=-\ln\left|\frac{\lambda_{2}}{\lambda_{1}}\right|,    
\end{equation}
where $\lambda_{1}$ and $\lambda_{2}$ are the two leading eigenvalues of the transfer matrix~\cite{schollwockDensitymatrixRenormalizationGroup2011a,rams_precise_2018-1}.
We plot $\xi^{-1}$ against the next inverse characteristic length related to the transfer-matrix spectrum, i.e. $\delta = -\ln|\lambda_{n+1}/\lambda_{n}|$ where $n>1$ is the smallest index such that $|\lambda_{n+1}|>|\lambda_n|$. For a critical state, both quantities are expected to approach zero together as the bond dimension is increased~\cite{rams_precise_2018-1,vanhecke_scaling_2019-1}. The observed behavior for the three phases that do not reach full order at $T=0$ is therefore consistent with a diverging correlation length and algebraically decaying correlations. %In particular, Phase 9, this is consistent with the mapping onto an effective triangular-lattice Ising antiferromagnet discussed above.

%%%%%%%%%%%%%%%%
\begin{figure*}
\begin{centering}
\includegraphics[width = 0.9\textwidth]{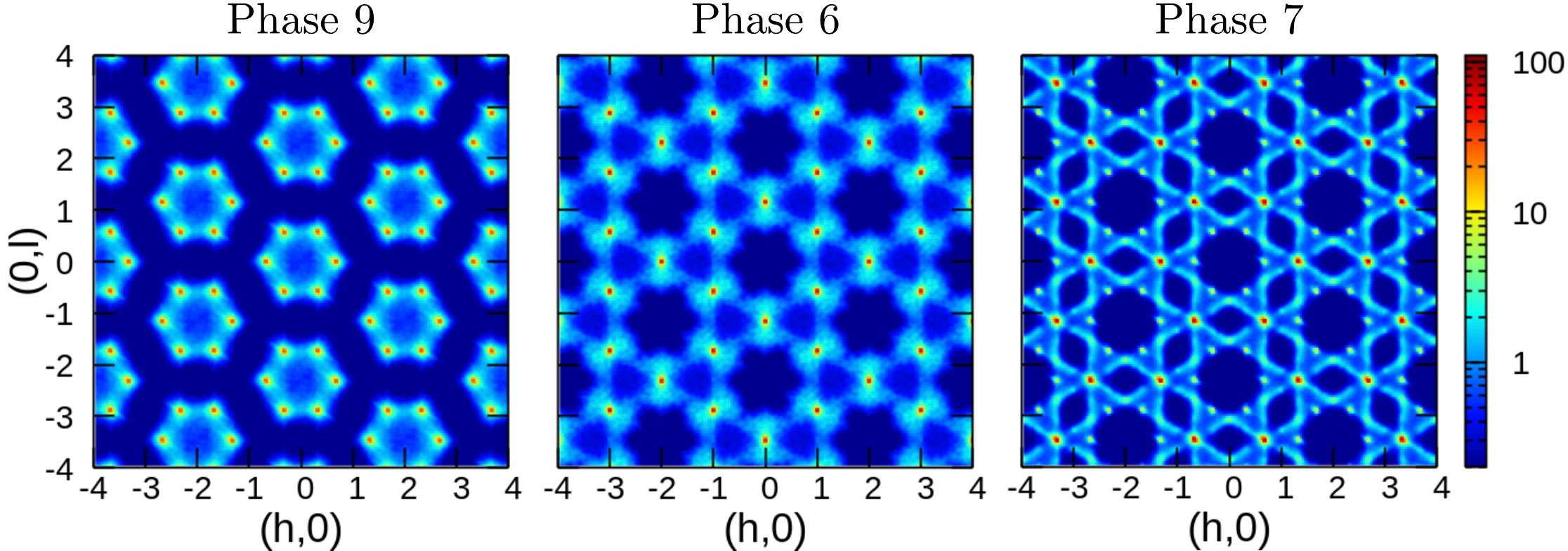}
\par\end{centering}
\caption{Magnetic structure factors for the three disordered phases found, one in the ferromagnetic case (Phase 9)
and two in the antiferromagnetic case (Phase 6 and Phase 7). 
The three panels show Bragg-like peaks, reflecting quasi long range order, coexisting with diffuse scattering. The patterns for each phase were produced for samples with $L=24$ at $T/|J_0|=0.01$, averaging over more than 30 statistically independent configurations in each case.
}
\label{fig:SQ}
\end{figure*}
%%%%%%%%%%%%%%%%

\subsection{Magnetic correlations}

To further characterize the macroscopically degenerate phases, we use Monte Carlo data to calculate the static structure factor $S(\mathbf{q})$, defined as the Fourier transform of the spin-spin correlation function,
\begin{equation}
    S(\boldsymbol{q})=\frac{1}{N}\sum_{i,j}\langle \sigma_{i} \sigma_{j}\rangle\,e^{i\boldsymbol{q}\cdot \boldsymbol{r}_{ij}},
\end{equation}
where $\langle \dots \rangle$ indicates thermal averaging. 

Fig.~\ref{fig:SQ} shows the structure factor for phases 9, 6, and 7 at very low temperature, and the same $J_i$ values used in Fig.~\ref{fig:EntropySF}. To make the coexistence of sharp peaks and diffuse scattering observed in all three cases more apparent, the intensity is displayed on a logarithmic scale. A common feature of Phase 9 (left) and Phase 7 (right) is the presence of sharp peaks at $\mathbf{q}=(2/3,0)$ and symmetry-related wave vectors. They are reminiscent of those found in the triangular-lattice Ising antiferromagnet, and signal real-space spin correlations with a periodicity of $3$ unit cells. On the other hand, there are clear differences between these patterns and that for the triangular-lattice structure factor. In particular, the peaks at $\mathbf{q}=(4/3,0)$ and symmetry related points are absent in Phase 9, whereas they are the dominant feature in Phase 7. This contrast highlights the three-site form-factor (sublattice) effect associated with the kagome lattice. The diffuse background is quite complex for Phase 7; remarkably, a closely related structure-factor pattern has previously been reported for a Heisenberg model on the kagome lattice~\cite{gen2022nematicity}. We return to this connection in Sec.~\ref{sec:dis}. In turn, Phase $6$ (center) shows a big central hexagon, rotated $90^\circ$ with respect to the other two phases. The value $\mathbf{q}=(0,\frac{2}{\sqrt{3}})\approx(0,1.155)$ points to correlations commensurate with the kagome lattice structure (between dense rows). 

While the narrow peaks observed on the three panels of Fig.~\ref{fig:SQ} are suggestive of Bragg peaks, we now argue that they only correspond to \emph{quasi}-long-range order. More precisely, we claim that the ground state phase, although macroscopically degenerate and with algebraically decaying correlations at long distance, is on the verge of ordering with a well-defined $q$-vector. This gives rise to narrow and intense peaks in the magnetic structure factor. 

Fig.~\ref{fig:algcorr} shows the spatial spin-spin correlations along selected directions of the kagome lattice, as obtained from CTMRG. As discussed above, the ground-state manifold of Phase 9 maps onto that of the triangular-lattice Ising antiferromagnet. One sublattice of kagome triangles spontaneously develops three aligned spins, and each such triangle acts as an effective Ising spin on a triangular lattice: $Q_{\triangle} := \sum_{i \in \triangle} \sigma_i$. 
This behavior is reminiscent of that discussed in Ref.~\onlinecite{Vanhecke2021}, but here, no additional zero-energy domain walls are present between domains of ferromagnetic up or ferromagnetic down triangles.
One might wonder whether the spontaneous symmetry breaking associated with the selection of up or down triangles leads to partial magnetic order (as is the case of the antiferromagnetic model with various types of third-neighbor couplings~\cite{dgyr-5y75,andrade_partial_2024-4}). However, since $Q_{\triangle}$ behaves as an effective Ising spin on the triangular lattice, its correlations are algebraically decaying. Indeed, Stephenson~\cite{Stephenson1970} showed that the spin-spin correlations along a row of the triangular lattice Ising antiferromagnet decay asymptotically as $\langle\sigma_{a}\sigma_{a+r}\rangle \propto \cos(2\pi r/3)/\sqrt{r}$. Consequently, the correlations along the sparse rows of the kagome lattice in phase 9 follow Stephenson's oscillatory asymptotic result for the triangular-lattice Ising antiferromagnet, after rescaling the distance $r$ to account for the spacing of the effective triangular lattice, as shown in Fig.~\ref{fig:algcorr}. Thus, the  absolute value of the disconnected correlations exhibits a clear power-law decay with exponent $\eta = 1/2$, as shown in the inset: the phase exhibits only quasi-long-range order.

Phase 6 does not show the same oscillations. Comparing with the structure factor in Fig.~\ref{fig:SQ}, this is compatible with a $q$ vector commensurate with the lattice spacing. Therefore, we only show the log-log plot, which highlights, again, a clear power-law decay with an exponent $\eta \sim 0.4 > 1/4$, suggesting a zero-temperature Villain-Stephenson critical point.

Finally, Phase 7 shares several features with Phase 9, most notably oscillatory correlations with an apparent power-law decay. There are, however, important differences. Unlike Phase 9, its ground-state manifold does not admit a direct mapping onto the triangular-lattice Ising antiferromagnet: additional local constraints produce more strongly correlated spin configurations, and we do not find a simple analytical form for the correlations. The right panel of Fig.~\ref{fig:algcorr} shows the correlations along a dense row of the kagome lattice. Their wave vector is consistent with the larger wave vector $\mathbf{q}=(4/3,0)$ associated with the corresponding peak in Fig.~\ref{fig:SQ}, while their envelope appears to decay slower than $1/\sqrt{r}$: the fitted exponent is $\eta \sim 0.32$. Together with the absence of a finite-temperature peak in the specific heat, the behavior shown in the inset provides strong evidence against conventional long-range order and is instead consistent with algebraically decaying correlations. Since $\eta > 1/4$, this again suggests a Villain-Stephenson critical point at zero temperature and an absence of finite-temperature critical phase. A more complete characterization of this unusual quasi-long-range ordered phase, and of the origin of its apparent critical behavior, remains an interesting direction for future work. 

%%%%%%%%%%%%%%%%%%%%%%%%%%%%%%%%%%%%%%%%%%%%%%%%%%%%%%%%
\begin{figure*}
\begin{centering}
\includegraphics[width = 0.98\textwidth]{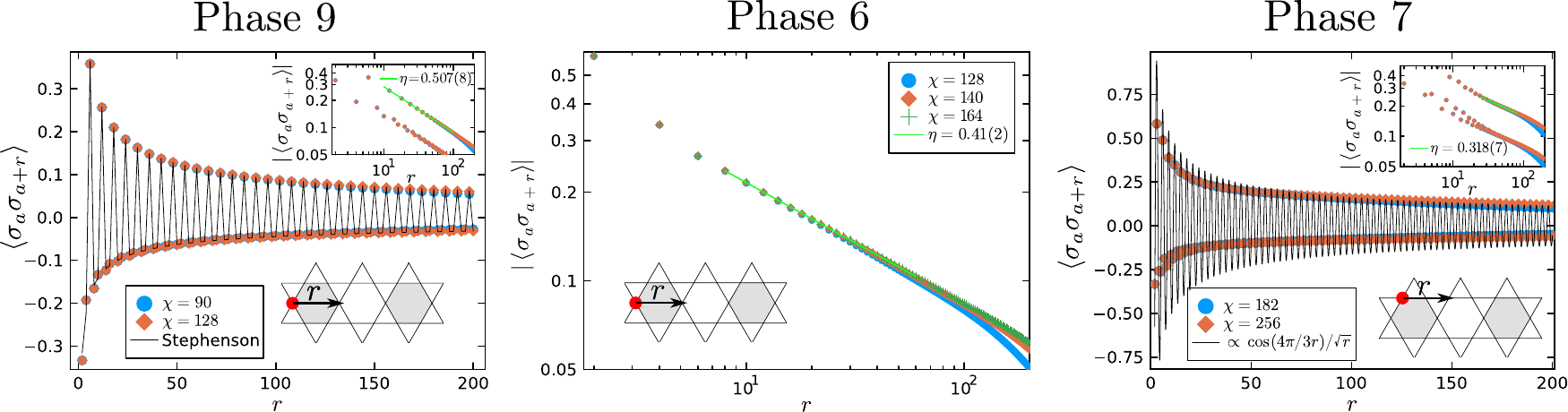}
\par\end{centering}
\caption{Evidence of algebraic decay of the (disconnected) spin-spin correlations in phases 9, 6 and 7 at zero temperature from CTMRG. In each panel, the position $a$ for the first spin is indicated by a red dot, and the direction along which the correlations are computed is indicated by an arrow. In all panels, $\chi$ indicates the bond dimensions: at long distances, the finite bond dimension induces a finite-size effect associated with the correlation length shown in Fig.~\ref{fig:convS}. This is well seen on the tail of the correlations in phase 6. In phase 9 and phase 7 panels, the black lines are not fits (see text). The insets in phase 9 and 7 and the main panel in phase 6 show evidence supporting the algebraic decay. The green lines indicate fits of the form $Ar^{-\eta}$ to the envelope of the correlations decay in the region where the two largest bond dimensions have converged. The couplings are the same as in Fig.~\ref{fig:convS}.}
\label{fig:algcorr}
\end{figure*}
%%%%%%%%%%%%%%%%%%%%%%%%%%%%%%%%

%%%%%%%%%%%%%%%%%%
\section{Implications for the magnetoelastic model \label{sec:back_to_magnetoelatics}}
%%%%%%%%%%%%%%%%%%

We can now revisit the magnetoelastic model that motivates this family of further neighbor Hamiltonians in light of the full $J_1-J_2-J_3$ phase diagram. As shown earlier, integrating out the lattice degrees of freedom in the magnetoelastic model generates trajectories through the phase diagram of the effective further-neighbor model, given by Eq.~\eqref{eq:kagome}, with the magnetoelastic coupling $\alpha$ acting as the control parameter (in fact, only $\alpha^2$ controls the trajectory). These trajectories are shown in Fig.~\ref{fig:PDkagome} as multicolored solid lines, with each color indicating the phase obtained from the full \textit{magnetoelastic} model of Ref.~\onlinecite{costa2026magnetic}. 

If we start from an undistorted ferromagnetic model, small values of $\alpha$ produce a reinforced effective coupling $J_1<0$ and small $J_2,\,J_3$ maintaining the system in a ferromagnetic ground state. This is indicated with the cyan segment in Fig.~\ref{fig:PDkagome}a). As $\alpha$ increases, the system reaches the first critical value,
\begin{equation}
\alpha_{c1} = \sqrt{\frac{2}{5}K_e},
\end{equation}
above which the ferromagnetic phase becomes unstable and transitions into the disordered Phase 9. This picture is in excellent agreement with the Monte Carlo simulations of the full magnetoelastic model in Ref.~\onlinecite{costa2026magnetic}, where the corresponding phase was referred to as \emph{self-frustrated}. The name reflects that this phase emerges from the spontaneous lattice distortions induced by the magnetoelastic coupling, which drive an otherwise unfrustrated ferromagnetic system into a frustrated state. The effective description also explains the absence of a nontrivial elastic contribution to the entropy in Ref.~\onlinecite{costa2026magnetic}: the spin configuration fixes the equilibrium positions about which the lattice undergoes harmonic vibrations (see Eq.~\eqref{eq:H_completedsquares}), leaving only the trivial harmonic contribution.

It is important to stress that Phase 9 has a true zero-point extensive entropy, while this is not exactly true for the full magnetoelastic model~\cite{costa2026magnetic}. Although connected by the integration of the structural degrees of freedom, the two models are not identical. After Eq.~(\ref{eq:distance_linear}), the full-spin model needs a further linearization process expressed by Eq. ~(\ref{eq:linearization}). 
As discussed in Ref.~\cite{costa2026magnetic}, this second linear approximation is good for temperatures down to $|J_0|/10^{3}$. 

As $\alpha$ is increased further, the system reaches a second critical value,
\begin{equation}\label{alphac2}
    \alpha_{c2}=\sqrt{K_e},
\end{equation}
at which the frustration is relieved by a transition into the checkerboard phase.

For an initially antiferromagnetic undistorted system, the evolution is simpler. Increasing $\alpha$ along the corresponding trajectory drives a single transition from the quasi-long-range ordered Phase 7, referred to as paramagnetic phase in Ref. \onlinecite{pili2019two}, into the  checkerboard phase  at
\begin{equation}\label{alphac3}
    \alpha_{c3}=\sqrt{\frac{K_e}{3}}\,.
\end{equation}
The magnetoelastic model remains in this phase with effective $J_1>0$ until $\alpha$ reaches the value $\sqrt{K_e}$, above which $J_1$ changes sign. This, however, does not imply a change of ground state: the system remains in the same checkerboard phase but on the ferromagnetic phase diagram of Fig.~\ref{fig:PDkagome} a).

%%%%%%%%%%%%%%%%%%%%%%%%%%%%%%%%%%
\section{Discussion} \label{sec:dis}
%%%%%%%%%%%%%%%%%%%%%%%%%%%%%%%%%%

Perhaps the most significant result of this work is the emergence of phases 6, 7, and 9 as macroscopically degenerate ground states with algebraically decaying spin correlations. These are features shared by very diverse frustrated systems~\cite{Stephenson1970,moessner1998low,fisher05_algebraicSL_PRB,yan2024classification,fennell2007pinch,costa2026magnetic,chern_magnetic_2012}). In the three macroscopically degenerate phases presented here, the quasi-long-range order is signaled by Bragg-like features in the spin structure factor.  Phase 7 is particularly interesting as it additionally exhibits (see right panel on Fig.~\ref{fig:SQ}) a left-over trace of the star-like patterns and half-moons observed on the fine-tuned $J_2 = J_3$ line ~\cite{mizoguchi2017clustering,mizoguchi_magnetic_2018-1,kiese_pinch-points_2023,yan2018half,yan2024classification} as well as a correspondence with a related Heisenberg model~\cite{gen2022nematicity}.

Let us first discuss the striking relationship between Phase 7 and the ``nematic phase'' reported for a magnetoelastic Heisenberg model on the kagome lattice~\cite{gen2022nematicity}, which exhibit remarkably similar structure factors. 
The origin of this correspondence becomes clear after recalling the consequences of integrating out the lattice degrees of freedom in the Heisenberg case. 
Both models treat the lattice degrees of freedom in the same way, using an Einstein site-phonon model and the same form of magnetoelastic coupling. 
After integration, in addition to the effective further-neighbor interactions, 
the resulting effective Hamiltonian in the Heisenberg case contains a biquadratic nearest-neighbor term that favors collinear spin configurations at sufficiently strong magnetoelastic coupling~\cite{gen2022nematicity}. 
Once colinearity is established~\footnote{Although the effect is the same as that observed in Refs.~\cite{moessner1998low,grison2020emergent}, colinearity is achieved through a very different mechanism in these cases (order by disorder). An even stronger equivalence between models exists for the dipolar correlations present in pyrochlores with a Gauss law as a local constraint, observed for all cases with spin dimensionality different from two~\cite{isakov04_dip_corr,henley_2005}.}, 
the remaining effective interactions reduce to the same form as those on our Ising model in Eq.~\eqref{eq:kagome}. 
The similarity of the structure factors of the two models can therefore be understood as a consequence of their common effective description in this regime.

The parallelism between the Heisenberg model in Ref.~\onlinecite{gen2022nematicity} and ours extends beyond Phase 7. Since a sufficiently strong magnetoelastic coupling drives the Heisenberg system into a collinear regime described by the same effective interactions as in our case, one expects the checkerboard phase to emerge in the Heisenberg model upon further increasing $\alpha$. This is indeed observed in Ref.~\onlinecite{gen2022nematicity}, where the related state is referred to as the \textit{N\'eel phase}. Remarkably, the transition occurs at the same critical value, $\alpha^2/K_e=1/3$, as in our Ising model [Eq.~\eqref{alphac3}].

At weaker magnetoelastic coupling, the near equivalence between both models necessarily breaks. In this regime the Heisenberg spins are not constrained to be collinear, and the coplanar $120^\circ$ state replaces Phase 7 of the Ising model as $\alpha\rightarrow0$. Ref.~\onlinecite{gen2022nematicity} considered only antiferromagnetic nearest-neighbor exchange. Our results nevertheless suggest a natural extension to the ferromagnetic Heisenberg case: if sufficiently strong magnetoelastic coupling enforces collinearity, an analogue of Phase 9 (the self frustrated phase in the coupled model) may also be stabilized.

As mentioned above, our model can also be connected with the hexamer spin liquid physics studied in Ref.~\onlinecite{mizoguchi2017clustering}, where the authors work with an Ising model on the kagome lattice which is the present $J_1-J_2-J_{3}$ model with $J_2 = J_{3}$, resulting in an effective attraction between emergent magnetic charges of the same sign. The low-temperature structure factors in this paper display half-moon features surrounding regions of destructive interference at the zone centers, which were associated with the formation of hexamer rings carrying identical magnetic charge. Phases 6 and 7 admit the $J_2 = J_{3}$ line as a boundary, and one might wonder whether the half-moon features extend into these phases. We find that closely related, although narrower and better defined, half-moon features appear in the structure factor of Phase 7, together with the peaks at $\mathbf{q}=(2/3,0)$ and symmetry-related wave vectors. While we have also detected the presence of same-charge hexamers in our spin configurations, the differences between the two structure factors can be naturally attributed to the different relative strengths of the effective further-neighbor interactions.

It is important to stress that we have been comparing the results obtained from a magnetoelastic model (where full Monte Carlo simulations, in which the spin and lattice degrees of freedom are considered on equal footing and sampled simultaneously) with its effective spin-only description. Despite its substantially lower computational cost, the effective model accurately reproduces the overall phase diagram and the critical values of $\alpha$ at which the phase transitions occur; it is then legitimate to ask if the full model can be totally dispensed with. We consider that some of the limitations of the spin only model become apparent when resolving the fine structure of the ground-state manifold. In particular, the linear approximation, Eq. (\ref{eq:linearization}), predicts an extensively degenerate self-frustrated phase, whereas simulations of the full magnetoelastic model show that this degeneracy is lifted at sufficiently low temperatures, selecting an ordered ground state~\cite{costa2026magnetic}. This difference indicates that higher-order terms in the expansion of the exchange interaction, neglected in the effective description, can become relevant when resolving the small energy differences between otherwise degenerate states.

This comparison also illustrates the complementary roles of the two descriptions. The full spin--lattice model retains the lattice degrees of freedom explicitly and therefore provides a natural framework for incorporating lattice effects and external perturbations, as illustrated in Ref.~\onlinecite{pili2022simple}. The effective model, by contrast, offers a considerably simpler and computationally less demanding description that captures the dominant physics and exposes its connection to frustrated spin models with further-neighbor interactions. This simplification comes at the cost of losing higher-order magnetoelastic effects. The self-frustrated phase provides a concrete example of this distinction.

%%%%%%%%%%%%%%%%%%%%%%%%%%%%%%%%%%
\section{Summary, conclusions, and perspectives} \label{sec:sum_conc}
%%%%%%%%%%%%%%%%%%%%%%%%%%%%%%%%%%

In this work, we have revisited a nearest-neighbor Ising model on a kagome lattice in which the magnetic ions are allowed to undergo elastic distortions. Integrating out the lattice degrees of freedom maps this magnetoelastic problem onto a purely magnetic model with further-neighbor interactions, generating a second-neighbor coupling $J_2$ and a coupling $J_{3}$ between one of the two inequivalent types of third neighbors. This mapping motivated us to study the more general phase diagram of the $J_1$-$J_2$-$J_{3}$ model.

Using complementary tensor network and Monte Carlo simulations, we determined its zero-temperature phase diagram with high numerical accuracy and characterized the degeneracy and entropy of its ground states. This allowed us to reconstruct, extend, and partially revise the phase diagram previously reported in the literature, revealing a rich interplay between ordered, partially ordered, and quasi-long-range ordered phases. In particular, we identified extensively degenerate phases, 6, 7, and 9, with finite zero-temperature entropy and signatures of algebraically decaying spin correlations, and clarified the nature and boundaries of several previously identified phases. The tensor-network approach proved especially valuable in this analysis, providing direct access to the ground-state degeneracy and spin correlations while avoiding the finite-size effects and slow low-temperature dynamics that can limit Monte Carlo simulations.

Our results also contribute to the broader picture of the remarkably rich phase behavior of kagome magnets with further-neighbor interactions~\cite{gong_global_2015-3,messio2011lattice,colbois2022partial,Colbois2021}. Particular attention has been devoted in the literature to the $J_{2}=J_{3}$ ($J_{3a}$) line, where characteristic half-moon patterns in the structure factor have been investigated in both Ising and Heisenberg models~\cite{mizoguchi2017clustering,mizoguchi_magnetic_2018-1,lugan_schwinger_2022-3,kiese_pinch-points_2023}, and where magnetization plateaus have also been reported~\cite{tokushuku_field-selective_2020,Colbois2021}. The present results show that unconventional behavior is not restricted to such fine-tuned conditions: several quasi-long-range ordered phases with nontrivial correlations can occupy extended regions of parameter space and, importantly, can arise naturally from a microscopic magnetoelastic mechanism. 

The finding of ground states with extensive entropy allowed us to establish connections with previous works on magnetoelastic models. As originally realized for decorated lattices~\cite{strevcka2012spin}, the distortion of a magnetic lattice without competing interactions into a frustrated one can now be understood within the framework of the pure spin model. In this way, the emergence of the \textit{self-frustrated} phase studied in Ref.~\cite{costa2026magnetic} springs from the competence between effective further-neighbor interactions; these interactions are able to stabilize Phase 9 in the ferromagnetic case, but not in the antiferromagnetic one~\cite{pili2019two}.
Also, we made a direct link between some of the phases detected in our Ising system and those found in the antiferromagnetic Heisenberg model of Ref.~\onlinecite{gen2022nematicity}. In a similar way, our results for $J_1<0$ may be used to speculate on the possible ground states that would be found in the ferromagnetic version of the Heisenberg model studied by Gen and collaborators \cite{gen2022nematicity}.

Several directions remain open.
At the level of the effective magnetic model, Phase 7 deserves further investigation, 
in particular to establish more precisely the nature of its algebraic correlations and their coexistence with local order. 
More broadly, the response of the magnetoelastic system to an applied magnetic field provides a natural extension of the present work 
and may reveal additional field-induced phases and 
magnetization plateaus~\footnote{The fact that magnetization plateaus exist away from the fine-tuned line was seen on phase 6 in Ref.~\onlinecite{Colbois2021}}. 
The present results may be relevant to frustrated kagome materials in which inequivalent bonds, as in breathing structures, modify the competition between magnetic interactions. Understanding how magnetoelastic coupling generates frustration and spontaneous bond inequivalence also provides a useful starting point for investigating frustrated magnets in which lattice distortions are also coupled to electric polarization, as required for magnetoelectric and multiferroic responses. 
Such extensions may help establish how the unconventional magnetic behavior uncovered here manifests itself in the broader spin--lattice response of kagome systems.

%%%%%%%%%%%%%%%%%%%%%%%%%%%%%%%%%%
\begin{acknowledgments}

We acknowledge the Agencia Nacional de Promoción Científica y Tecnológica (ANPCyT) Argentina, for awarding grants  PICT 2022-11-00046 and PICT 2022-11-00100, which, although financed by the Inter-American Development Bank (IDB), were later discontinued following changes in national funding policy. SAG is a Royal Society Wolfson Visiting Fellow at Oxford University funded by the Wolfson Foundation and Royal Society. The authors are partially supported by Consejo Nacional de Investigaciones Cient\'ificas y T\'ecnicas (CONICET), Argentina. 
J.C. acknowledges financial support from CNRS (BONAC) for the computational resources used in this study. The tensor network calculations were performed using the Julia packages PEPSKit.jl~\cite{Brehmer2026PEPSKit} and TensorKit.jl~\cite{tensorkit}.

\end{acknowledgments}

\appendix

\section{Tensor network construction} \label{app:TN}
In this appendix, we provide a reminder of the construction of tensor networks for evaluating partition functions of 2D classical frustrated Ising models. This appendix reproduces some of the main observations of Refs.~\cite{Orus2008, Orus2009, Nishino1995, Vanhecke2021, Colbois2021, colbois2022partial}.
\subsection{Partition function of 2D Ising models as tensor networks}

For a generalized Ising model with a finite range of couplings with pair interactions and a given temperature, it is clear that at least one representation of the partition function as a single layer two-dimensional TN exists, with bond dimension $d = 2$ (see for instance Ref.~\citep{Orus2009}). Its structure corresponds to the graph of the model including all couplings (e.g.,  a planar graph for the nearest-neighbor square lattice Ising model, and a graph with crossings for the next-nearest-neighbor Ising model) . It is built by associating to each coupling a two-leg tensor representing the Boltzmann weight associated with this coupling, $t_{\sigma_{i}\sigma_{j}} = \exp(-\beta J_{i,j} \sigma_i \sigma_{j})$, and to each site of the lattice a generalization of a Kronecker-$\delta$ with as many legs as there are couplings. The latter ensures that, upon contraction, the only non-zero entries are those where all Boltzmann weight tensors are evaluated for the same spin configuration. Since the contraction of a tensor network corresponds to the sum over all virtual degrees of freedom of the product of the tensor entries, the partition function is the exact contraction of this tensor network. However, except in exactly solvable cases, it is obtained approximately using standard TN algorithms. 
While this construction is conceptually simple to understand, it is not always practical because the low temperature limit of the local tensor is often ill-conditioned in the case of frustrated models, creating issues in the approximate contraction algorithm~\cite{Vanhecke2021, 2022-Colbois,liu_tropical_2021,tang_matrix_2025}.
Both these issues are addressed by the alternative construction below. 

\subsection{Local tensor construction for frustrated Ising models}
A given partition function admits various 2D TN representations. While in the exact contraction they would be equivalent, the convergence of the (approximate) contraction algorithms depends on the exact choice of the TN representation. We briefly describe a simplified version of the TN construction for frustrated Ising models introduced in Ref.~\onlinecite{Vanhecke2021}. 

Such a construction is typically necessary when the system is frustrated, namely in the natural splitting of the Hamiltonian into bonds, each bond energy cannot be simultaneously minimized. This prevents having both a well-normalized partition function at zero temperature and a well-defined Boltzmann-weight bond tensor in the zero-temperature limit~\cite{Vanhecke2021}.
In simple examples of nearest-neighbor frustrated models on the kagome or triangular lattice, or in the next-nearest-neighbor model on the square lattice, it is clear that a well-normalized construction is to associate a Boltzmann weight with each plaquette of the model, such that the Hamiltonians on the plaquettes can be simultaneously minimized and the Boltzmann weight on the plaquette can be normalized such that it goes to either $1$ or $0$ in the zero temperature limit. 

The following procedure systematizes this construction to farther-neighbor models. The central idea is to select a cluster shape $\mathcal{C}$, for instance a kagome star on the lattice, and to build the tensor network using the Boltzmann weights on each of these (overlapping) clusters. To this effect, we rewrite the total Hamiltonian $H$ as a sum of local Hamiltonians $h^{\boldsymbol{\gamma}}_{\mathcal{C}}$ on overlapping clusters translated over the lattice. The parameters $\boldsymbol{\gamma}$ specify how a given coupling is split between two or more overlapping clusters which share it. For any configuration $\boldsymbol{\sigma}|_{\mathcal{C}}$ of spins on the cluster, there is an associated Boltzmann weight  $e^{- \beta h^{\gamma}_{\mathcal{C}}(\boldsymbol{\sigma}|_{\mathcal{C}})}$.  For a given choice of $\boldsymbol{\gamma}$ (see below), the partition function is a translation invariant tensor network which associates to each cluster a tensor $T_{\boldsymbol{\sigma}|_{\mathcal{C}}}= e^{- \beta h^{\gamma}_{\mathcal{C}}(\boldsymbol{\sigma}|_{\mathcal{C}})}$ with as many (virtual) legs as there are sites on the cluster. Two tensors corresponding to overlapping clusters are connected through the legs that correspond to their shared sites, ensuring that only compatible spin configurations between the two clusters are counted.

\begin{figure}
\centering
\includegraphics[width=0.95\linewidth]{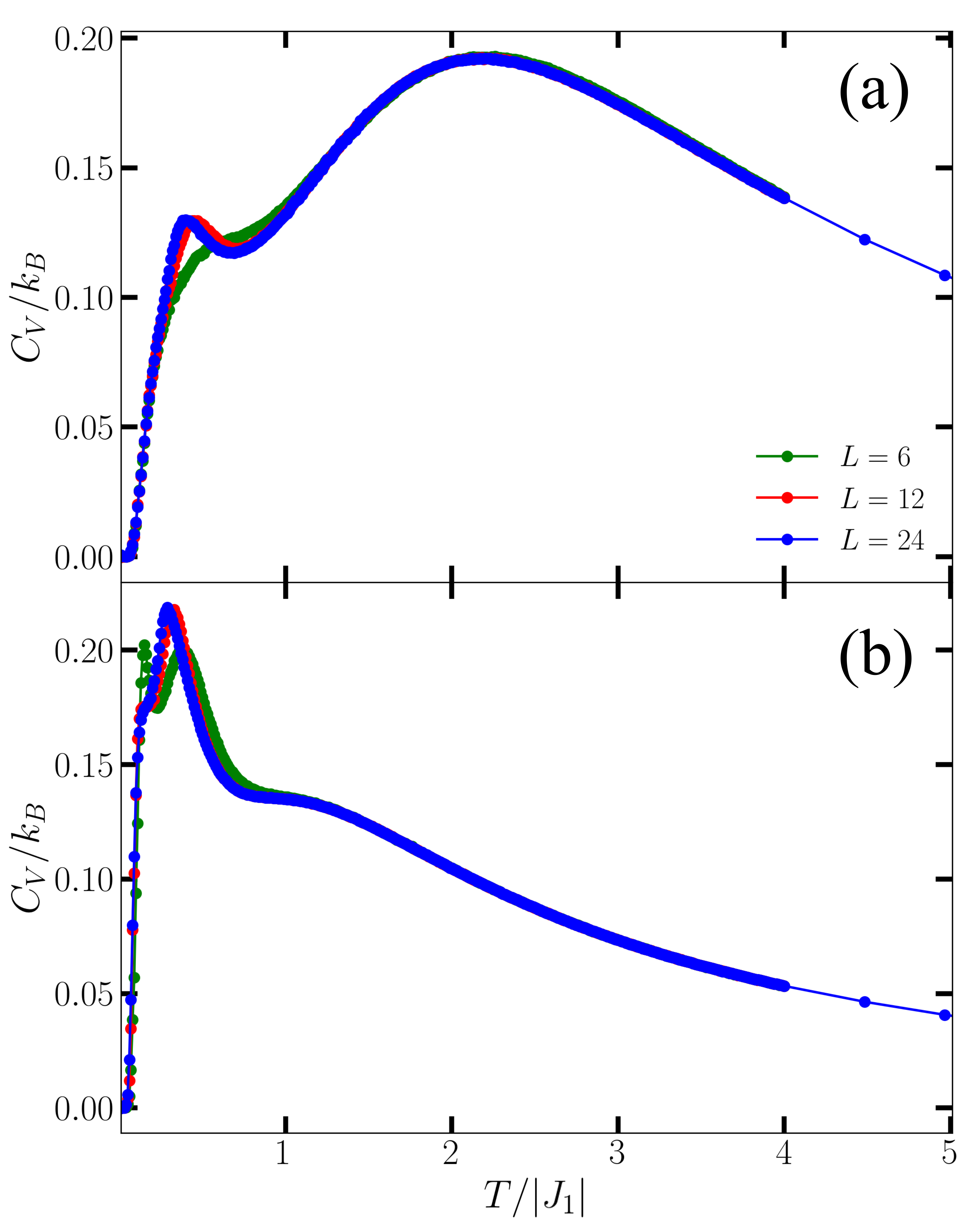}
\caption{Specific heat per spin as a function of temperature, for Phase 6 (a) and Phase 7 (b), measured at different system sizes $L$. Although there is some change for low $L$, the peak height and width seems to saturate for moderately large sizes.}
\label{fig:FSE_phases}
\end{figure}

The weights $\boldsymbol{\gamma}$ can be selected as follows, given a certain Hamiltonian $H$, a cluster, and values of the coupling. First, that for each coupling, the weights must sum to one. Then, the idea is to test whether the cluster relaxed the frustration. For any choice, $E_{\rm LB}^{\boldsymbol{\gamma}} := \min_{\boldsymbol{\sigma}|_{\mathcal{C}}}h^{\alpha}_{\mathcal{C}}(\boldsymbol{\sigma}|_{\mathcal{C}})$ is a \emph{lower bound} for the ground state energy per cluster, which can therefore be optimized by maximizing $E_{\rm LB}^{\boldsymbol{\gamma}}$ over $\boldsymbol{\gamma}$ under constraint. This can be tried systematically as it can be written as a linear program~\cite{Vanhecke2021}. The linear program is significantly simplified by optimizing only over weights $\boldsymbol{\gamma}$ that respect lattice symmetries~\cite{2022-Colbois}. 

In practice, for the kagome $J_1-J_2-J_{3}$ model, a single-star cluster suffices. Using symmetries, there remains only two free weights: one which specifies the splitting of the nearest-neighbor bonds, and one for the splitting of next-nearest-neighbor bonds. 

Finally, we make two remarks on the zero temperature limit. First, the construction provides either a lower bound or an exact result for the ground state energy. The result is exact if the local configurations that minimize the Hamiltonian on the cluster can be matched consistently to form at least one global spin configuration on the lattice~\cite{Huang2016}. For instance, in the triangular lattice Ising antiferromagnet, we would get a lower bound by using 2-sites links as clusters, but an exact results by using triangles as clusters.  
Second, in that limit, the local tensor simplifies significantly: 
\begin{equation}
    T_{\boldsymbol{\sigma}|_{\mathcal{C}}}= \begin{cases}
        1 \quad \text{if } h^{\boldsymbol{\gamma}*}(\boldsymbol{\sigma}|_{\mathcal{C}}) = E_{\rm LB}^{\boldsymbol{\gamma}*}\\
        0 \quad \text{otherwise}
    \end{cases}, 
\end{equation}
where $\boldsymbol{\gamma}*$ is an optimal set of weights.

\section{Finite size effects on the disordered phases 
\label{app:finite_size_effects}
}

In the main text, we noted that the specific heat of phases 6 and 7 shows no finite-size behavior indicative of a first- or second-order phase transition. To illustrate this point, Fig.~\ref{fig:FSE_phases} shows the specific heat per spin as a function of temperature for system sizes $L=6$, $12$, and $24$. Although the shape of the curves exhibits some size dependence, particularly for the smallest system, the evolution with $L$ differs markedly from that expected near a conventional first- or second-order phase transition. In particular, the height of the specific-heat peaks appears to saturate as the system size increases, providing no evidence for a divergent thermodynamic anomaly.

%%%%%%%%%%%%%%%%%%%%%%%%%%%%%%
\bibliographystyle{apsrev4-2}
\bibliography{postdoc}% Produces the bibliography via BibTeX.

\end{document}